УДК 681.51

# Адаптивный наблюдатель переменных состояния нелинейной нестационарной системы с неизвестными постоянными параметрами

**Авторы:**

Козачёк Ольга Андреевна, аспирант, oakozachek@itmo.ru, федеральное государственное автономное образовательное учреждение высшего образования «Национальный исследовательский университет ИТМО», Кронверкский пр., д. 49, Санкт-Петербург

Алексей Алексеевич Бобцов, д-р техн. наук, проф., bobtsov@mail.ru, федеральное государственное автономное образовательное учреждение высшего образования «Национальный исследовательский университет ИТМО», Кронверкский пр., д. 49, Санкт-Петербург

Николаев Николай Анатольевич канд. техн. наук, доц., nikona@yandex.ru, федеральное государственное автономное образовательное учреждение высшего образования «Национальный исследовательский университет ИТМО», Кронверкский пр., д. 49, Санкт-Петербург

**Аннотация:** В статье предложен адаптивный наблюдатель вектора состояния нелинейной нестационарной системы по измерениям выходной переменной. Задача решена в предположении, что матрица (вектор) управления и нелинейный компонент уравнения состояния системы содержат неизвестные постоянные параметры. При разработке адаптивного наблюдателя был применен метод GPEBO (generalized parameter estimation based observer), также известный, как обобщенный наблюдатель, основанный на оценке параметров, который был предложен в работе [1]. При синтезе наблюдателя проводится предварительная параметризация исходной нелинейной системы. Затем полученная система приводится к линейной регрессионной модели. На следующем этапе неизвестные постоянные параметры регрессии оцениваются с помощью метода наименьших квадратов с фактором забывания (forgetting factor) [2, 3]. В статье предлагается развитие результата, предложенного авторами в работе [4]. В [4] была рассмотрена линейная нестационарная система, содержащая неизвестные параметры в матрице (векторе) управления. Данная работа является расширением результата, полученного в [4] на случай, когда уравнение состояния системы содержит частично неизвестную нелинейность.

**Ключевые слова:** адаптивный наблюдатель, нелинейная система, нестационарная система, линейная регрессионная модель, идентификация параметров

# Adaptive state observer for nonlinear time-variant system with unknown constant parameters

**Authors:**

Bobtsov Alexey Alexeevich, PhD, professor, bobtsov@mail.ru, ITMO University, Kronverkskiy pr., d. 49, Saint-Petersburg

Nikolaev Nikolay Anatolievich, PhD, docent, nikona@yandex.ru, ITMO University, Kronverkskiy pr., d. 49, Saint-Petersburg

Kozachek Olga Andreevna, postgraduate, oakozachek@itmo.ru, ITMO University, Kronverkskiy pr., d. 49, Saint-Petersburg

**Abstract:** An adaptive state vector observer for a nonlinear time variant system with a measurable output is developed in this paper. The task is solved under assumption that the input matrix (vector) and the nonlinear component in the state equation of the system contain unknown constant parameters. The developed adaptive observer is based on generalized parameter estimation based observer (GPEBO approach), which was proposed in [1]. During observer synthesis preliminary parametrization of the initial nonlinear system was made. After that the obtained system is transformed into linear regression model. The next step of the algorithm is estimation of the unknown linear regression parameters. The least square method with forgetting factor [2, 3] is used for that. In the paper [4] linear time-variant system with unknown parameters in the state matrix and the input matrix (vector) was considered. The current work is to develop the result obtained in [4] for the case when the state equation of the system contains an unknown nonlinear component.

This work was supported by Russian Science Foundation, project no. 22-21-00499.

**Введение**

Получение информации о состоянии системы является важной задачей при управлении динамическими системами. Для этого могут использоваться первичные измерительные преобразователи (датчики). Однако не всегда весь вектор состояния объекта доступен прямым измерениям. В случаях, когда невозможно разместить набор измерительных средств, достаточный для измерения всего вектора состояния, для оценки неизвестных переменных применяются наблюдатели.

Методы синтеза наблюдателей состояния линейных динамических систем с постоянными параметрами известны и достаточно эффективны [5, 6]. Однако интерес исследователей к проблеме синтеза наблюдателей для линейных систем не угасает. Об этом свидетельствует публикационная активность в изданиях, посвященных проблемам анализа и синтеза систем автоматического управления. В частности, в [6] была рассмотрен синтез оптимальных эллипсоидных наблюдателей и алгоритмов, которые позволяют обеспечить оптимальные эллипсоидные оценки вектора состояния системы и неизвестных параметров.

Проблема разработки алгоритмов наблюдения переменных состояния нелинейных систем на данный момент изучена меньше. По этой причине интерес к исследованию нелинейных систем в научном сообществе сохраняется [7, 8].

Важным аспектом в задаче построения наблюдателей является тот факт, что объект не всегда может быть описан моделью с постоянными параметрами. В некоторых случаях параметры системы изменяются со временем. Это может быть обусловлено внутренними и внешними факторами. Среди таких факторов, например, изменение параметров вследствие старения элементов системы, воздействие экстремальных температур, изменение массо-габритных параметров в ходе эксплуатации. В связи с этим поведение сложных динамичных систем может быть описано более точно с помощью математических моделей, содержащих нестационарные параметры. По этой причине исследования, посвященные проблеме синтеза наблюдателей для нестационарных систем, в настоящее время имеет большую актуальность.

При построении наблюдателей применяются различные подходы. Одним из подходов является сведение исходной модели объекта к линейной регрессионной модели (см., например, [9, 10]) с дальнейшей идентификацией неизвестных параметров модели.

Алгоритмы оценки вектора состояния нелинейной нестационарной системы могут использоваться не только при синтезе законов управления. Они также имеют и самостоятельное значение. К примеру, данные алгоритмы могут применяться при разработке средств контроля технического состояния [11–12].

При решении задач, связанных с синтезом наблюдателей неизвестных переменных состояния нестационарных систем, исследователи вводят в отношении матриц описания объекта различные допущения и предположения. В качестве примера можно рассмотреть работу [13], где предполагается, что матрица состояния задана в канонической форме. Кроме того, в статье [14] предполагается, что матрица состояния может быть представлена в виде суммы, где одно слагаемое известно, а второе состоит из неизвестных постоянных параметров. В статье в первую очередь оцениваются неизвестные параметры, а затем на основе полученных оценок синтезируется наблюдатель неизвестного вектора состояния системы.

В настоящей работе предложено развитие результатов, ранее опубликованных авторами в работе [4], где был представлен алгоритм оценки вектора состояния линейной нестационарной системы. Рассмотренная система содержит неизвестные параметры в матрице состояния и матрице (векторе) управления. В настоящей работе подход, предложенный в [4], развивается на случай, когда система является нелинейной. Предполагается, что нелинейный компонент, содержащийся в уравнении состояния системы, частично неизвестен.

**Постановка задачи**

Рассматривается нелинейная нестационарная система с одним входом и одним выходом (SISO) вида:

$$\dot{\mathbf{x}}(t) = \mathbf{A}(t)\mathbf{x}(t) + \mathbf{k}\mathbf{C}^{\mathrm{T}}(t)\mathbf{x}(t) + \mathbf{b}u(t) + \mathbf{w}(y,t), \ \mathbf{x}(0) = \mathbf{x}_0 \in \mathbb{R}^n, t \geq 0,$$
$$y(t) = \mathbf{C}^{\mathrm{T}}(t)\mathbf{x}(t),$$
(1)

где $\mathbf{x}(t) \in \mathbb{R}^n$ – неизвестный вектор состояния, $u(t) \in \mathbb{R}$ – известный входной сигнал, $y(t) \in \mathbb{R}$ – измеряемый выходной сигнал, матрицы $\mathbf{A}(t) \in \mathbb{R}^{n \times n}$, $\mathbf{C}^{\mathrm{T}}(t) \in \mathbb{R}^n$ являются известными и ограниченными матрицами с нестационарными параметрами, $\mathbf{k} \in \mathbb{R}^n$ и $\mathbf{b} \in \mathbb{R}^n$ – постоянны и неизвестны, $\mathbf{w}(y,t)$ – частично неизвестная нелинейная вектор-функция.

В отношении рассматриваемой системы при решении поставленной задачи были приняты следующие допущения.

*Допущение 1.* Нелинейная вектор-функция $\mathbf{w}(y,t)$ может быть представлена в виде:

$$\mathbf{w}(y,t) = \mathbf{m}f(y(t)),$$

где $f(y(t))$ – известная нелинейная функция, а $\mathbf{m} \in \mathbb{R}^n$ – вектор неизвестных постоянных параметров.

*Допущение 2.* Предполагается, что траектории входа и состояния ограничены.

*Допущение 3.* Пара матриц $\mathbf{A}(t)$ и $\mathbf{C}^{\mathrm{T}}(t)$ обнаруживаема. Это означает, что существует матрица обратной связи $\mathbf{L}(t)$ такая, что автономная система

$$\dot{\mathbf{x}}(t) = [\mathbf{A}(t) - \mathbf{L}(t)\mathbf{C}^{\mathrm{T}}(t)]\mathbf{x}(t),$$

асимптотически устойчива.

*Допущение 4.* Автономная система $\dot{\mathbf{x}}(t) = \mathbf{A}_0(t)\mathbf{x}(t)$, где $\mathbf{A}_0(t) = \mathbf{A}(t) - \mathbf{L}(t)\mathbf{C}^{\mathrm{T}}(t)$, является равномерно устойчивой (uniformly stable). То есть её фундаментальная матрица удовлетворяет условию (см. Теорему 6.4 [15]):

$$\|\mathbf{\Phi}_{\mathbf{A}_0}(t,\tau)\| \leq c_1, \forall\, t \geq \tau \geq 0.$$

Введённые допущения являются типовыми (см., например, [15–17]).

Для системы (1) ставится задача синтеза адаптивного наблюдателя вида:

$$\dot{\boldsymbol{\chi}}(t) = \mathbf{F}\big(\boldsymbol{\chi}(t), u(t), y(t)\big),$$

$$\begin{bmatrix} \hat{\mathbf{x}}(t) \\ \hat{\mathbf{k}}(t) \\ \hat{\mathbf{b}}(t) \\ \hat{\mathbf{m}}(t) \end{bmatrix} = \mathbf{S}\big(\boldsymbol{\chi}(t), u(t), y(t)\big),$$

где $\boldsymbol{\chi}(t) \in \mathbb{R}^{n_\chi}$ такое, что все сигналы ограничены. Адаптивный наблюдатель должен обеспечивать сходимость оценок переменных состояния и оценок постоянных неизвестных параметров к реальным значениям:

$$\hat{\mathbf{x}}(t) = \mathbf{x}(t),\ \hat{\mathbf{k}}(t) = \mathbf{k},\ \hat{\mathbf{b}}(t) = \mathbf{b}, \hat{\mathbf{m}}(t) = \mathbf{m},$$

для всех $\mathbf{x}_0 \in \mathbb{R}^n, \boldsymbol{\chi}(t) \in \mathbb{R}^{n_\chi}$.

## Основной результат

На первом этапе данной работы производится параметризация исходной системы с целью получения статической линейной регрессионной модели. Таким образом, основной задачей становится идентификация неизвестных постоянных параметров линейной статической регрессионной модели. В дальнейшем на основе полученных параметров можно будет восстановить компоненты вектора состояния. На втором шаге работы производится оценка неизвестных постоянных параметров линейной регрессионной модели. Для решения данной задачи существует множество различных методов. Выбор метода зависит от условий возбуждения, накладываемых на регрессор, см., например, [10, 18, 19].

**Теорема 1.** Рассмотрим динамическую систему вида

$$\dot{\boldsymbol{\xi}}(t) = \mathbf{A}_0(t)\boldsymbol{\xi}(t) + \mathbf{L}(t)y(t), \quad \boldsymbol{\xi}(0) = \mathbf{0}_{n\times 1}, \quad (2)$$

$$\dot{\boldsymbol{\eta}}(t) = \mathbf{A}_0(t)\boldsymbol{\eta}(t) + \mathbf{I}y(t), \quad \boldsymbol{\eta}(0) = \mathbf{0}_{n\times n}, \quad (3)$$

$$\dot{\boldsymbol{\zeta}}(t) = \mathbf{A}_0(t)\boldsymbol{\zeta}(t) + \mathbf{I}u(t), \quad \boldsymbol{\zeta}(0) = \mathbf{0}_{n\times n}, \quad (4)$$

$$\dot{\boldsymbol{\rho}} = \mathbf{A}_0(t)\boldsymbol{\rho}(t) + \mathbf{I}f(y), \quad \boldsymbol{\rho}(0) = \mathbf{0}_{n\times n}, \quad (5)$$

$$\dot{\boldsymbol{\Phi}}(t) = \mathbf{A}_0(t)\boldsymbol{\Phi}(t), \quad \boldsymbol{\Phi}(0) = \mathbf{I}_{n\times n}, \quad (6)$$

где матрицы $\mathbf{A}_0(t)$ и $\mathbf{L}(t)$ удовлетворяют допущениям 3…5, а $\mathbf{I}$ – единичная матрица соответствующей размерности.

Таким образом, исходную динамическую систему (1) можно преобразовать к линейной регрессионной модели в новых переменных вида

$$z(t) = \boldsymbol{\Psi}(t)\boldsymbol{\Theta}, \quad (7)$$

где сигнал $z(t) = y(t) - \mathbf{C}^\mathrm{T}(t)\boldsymbol{\xi}(t)$ измеряется, $\boldsymbol{\Psi}(t) = [-\mathbf{C}^\mathrm{T}(t)\boldsymbol{\Phi}(t) \quad \mathbf{C}^\mathrm{T}(t)\boldsymbol{\eta}(t) \quad \mathbf{C}^\mathrm{T}(t)\boldsymbol{\zeta}(t) \quad \mathbf{C}^\mathrm{T}(t)\boldsymbol{\rho}(t)]$ – вектор известных функций, $\boldsymbol{\Theta} = [\boldsymbol{\theta} \quad \mathbf{k} \quad \mathbf{b} \quad \mathbf{m}]^\mathrm{T} = [\Theta_1 \quad \Theta_2 \quad \Theta_3 \quad \Theta_4 \quad \Theta_5 \quad \Theta_6 \quad \Theta_7 \quad \Theta_8]^\mathrm{T}$ – вектор неизвестных постоянных параметров.

*Доказательство.*

Рассмотрим уравнение ошибки вида

$$\mathbf{e}(t) = \boldsymbol{\xi}(t) + \boldsymbol{\eta}(t)\mathbf{k} + \boldsymbol{\zeta}(t)\mathbf{b} + \boldsymbol{\rho}(t)\mathbf{m} - \mathbf{x}(t). \quad (8)$$

Производная ошибки $\dot{\mathbf{e}}(t)$ имеет вид:

$$\dot{\mathbf{e}}(t) = \dot{\boldsymbol{\xi}}(t) + \dot{\boldsymbol{\eta}}(t)\mathbf{k} + \dot{\boldsymbol{\zeta}}(t)\mathbf{b} + \dot{\boldsymbol{\rho}}(e)\mathbf{m} - \dot{\mathbf{x}}(t) = \mathbf{A}_0(t)\boldsymbol{\xi}(t) + \mathbf{L}(t)y(t) +$$
$$+ \mathbf{A}_0(t)\boldsymbol{\eta}(t)k + \mathbf{I}y(t)\mathbf{k} + \mathbf{A}_0(t)\boldsymbol{\zeta}(t)\mathbf{b} + \mathbf{I}u(t)\mathbf{b} + \mathbf{A}_0(t)\boldsymbol{\rho}(t)\mathbf{m} + \mathbf{I}f(y)\mathbf{m} -$$
$$- \left(\mathbf{A}_0(t) + \mathbf{L}(t)\mathbf{C}^T(t)\right)\mathbf{x}(t) - \mathbf{k}\mathbf{C}^\mathrm{T}(t)\mathbf{x}(t) - \mathbf{b}u(t) - \mathbf{m}f(y) =$$
$$= \mathbf{A}_0(t)(\boldsymbol{\xi}(t) + \boldsymbol{\eta}(t)\mathbf{k} + \boldsymbol{\zeta}(t)\mathbf{b} + \boldsymbol{\rho}(t)\mathbf{m}) - \mathbf{A}_0(t)\mathbf{x}(t) = \mathbf{A}_0\mathbf{e}(t)$$
$$\dot{\mathbf{e}}(t) = \mathbf{A}_0\mathbf{e}(t).$$

Решением полученного дифференциального уравнения является функция:

$$\mathbf{e}(t) = \boldsymbol{\Phi}(t)\boldsymbol{\theta}, \quad (9)$$

где $\boldsymbol{\theta} = \mathbf{e}(0)$ – начальные условия вектора $\mathbf{e}(t)$. При нулевых начальных условиях динамической системы (2)…(6):

$$\mathbf{e}(0) = -\mathbf{x}(0).$$

Выполнив подстановку (9) в (8), получим:

$$\mathbf{x}(t) - \boldsymbol{\xi}(t) = \boldsymbol{\eta}(t)\mathbf{k} + \boldsymbol{\zeta}(t)\mathbf{b} + \boldsymbol{\rho}(t)\mathbf{m} - \boldsymbol{\Phi}(t)\boldsymbol{\theta}. \qquad (10)$$

Преобразуем полученное выражение, умножив обе части равенства (10) на $\mathbf{C}^{\mathrm{T}}(t)$. Полученное уравнение является линейной регрессионной моделью вида (7).

Для оценки неизвестных параметров линейной регрессии могут быть применены различные подходы. Среди этих подходов, например, градиентный алгоритм идентификации [20], метод динамического расширения регрессора и смешивания [18, 19] и др. Стоит отметить, что большинство методов может быть применено только в случае, если регрессия удовлетворяет условию неисчезающего возбуждения:

$$\alpha_2 \mathbf{I} \leq \int_{t_0}^{t_0+\delta} \boldsymbol{\Psi}(\tau)\boldsymbol{\Psi}^T(\tau)d\tau \leq \alpha_1 \mathbf{I}, \text{ для всех } t_0 > 0,$$

где $\alpha_1$, $\alpha_2$ и $\delta$ – положительные константы.

В данной работе для оценки неизвестных параметров линейной регрессионной модели (7) был выбран метод наименьших квадратов с фактором забывания (forgetting factor) [2, 3]:

$$\dot{\widehat{\boldsymbol{\Theta}}} = \gamma \mathbf{F}(t)\boldsymbol{\Psi}^{\mathrm{T}}(t)\bigl(z(t) - \boldsymbol{\Psi}(t)\widehat{\boldsymbol{\Theta}}\bigr),$$

$$\dot{\mathbf{F}} = \begin{cases} -\gamma \mathbf{F}(t)\boldsymbol{\Psi}^{\mathrm{T}}(t)\boldsymbol{\Psi}(t)\mathbf{F}(t) + \beta \mathbf{F}(t), & \text{если } \|\mathbf{F}(t)\| \leq \mathrm{M} \\ \mathbf{0} & \text{если } \|\mathbf{F}(t)\| > \mathrm{M}' \end{cases}$$

где $\mathbf{F}(0) = \frac{1}{f_0}\mathbf{I}$, где $\mathbf{I}$ – единичная матрица, $\gamma > 0$, $\beta > 0$, $f_0 \geq 0$, $\mathrm{M} > 0$ – настраиваемые параметры, а вектор $\widehat{\boldsymbol{\Theta}} = [\widehat{\boldsymbol{\theta}} \quad \widehat{\mathbf{k}} \quad \widehat{\mathbf{b}} \quad \widehat{\mathbf{m}}]^{\mathrm{T}} = [\widehat{\Theta}_1 \quad \widehat{\Theta}_2 \quad \widehat{\Theta}_3 \quad \widehat{\Theta}_4 \quad \widehat{\Theta}_5 \quad \widehat{\Theta}_6 \quad \widehat{\Theta}_7 \quad \widehat{\Theta}_8]^{\mathrm{T}}$ – оценка вектора неизвестных параметров $\boldsymbol{\Theta}$.

После получения оценок неизвестных постоянных параметров системы, оценку вектора состояния можно найти с помощью (10) в виде:

$$\hat{\mathbf{x}}(t) = \boldsymbol{\xi}(t) - \boldsymbol{\Phi}(t)\begin{bmatrix}\hat{\Theta}_1\\\hat{\Theta}_2\end{bmatrix} + \boldsymbol{\eta}(t)\begin{bmatrix}\hat{\Theta}_3\\\hat{\Theta}_4\end{bmatrix} + \boldsymbol{\zeta}(t)\begin{bmatrix}\hat{\Theta}_5\\\hat{\Theta}_6\end{bmatrix} + \boldsymbol{\rho}(t)\begin{bmatrix}\hat{\Theta}_7\\\hat{\Theta}_8\end{bmatrix}.$$

### Результаты моделирования

При моделировании алгоритма использовалась система (1) со следующими параметрами:

$$\mathbf{A}(t) = \begin{bmatrix}2 - \sin t & 1\\-8 + \cos(t) & 0\end{bmatrix}, \mathbf{b} = \begin{bmatrix}1\\2\end{bmatrix}, \mathbf{k} = \begin{bmatrix}-1\\3\end{bmatrix}, \mathbf{C}(t) = \begin{bmatrix}1\\0\end{bmatrix},$$

$$\mathbf{w}(y,t) = \mathbf{m}\sin(y),$$

где $\mathbf{m} = \begin{bmatrix}-4\\4\end{bmatrix}$.

Используя вектор $\mathbf{L}(t) = \begin{bmatrix}2 - \sin(t)\\1 + \cos(t)\end{bmatrix}$, имеем

$$\mathbf{A}_0(t) = \begin{bmatrix}0 & 1\\-9 & 0\end{bmatrix}.$$

Начальные условия вектора состояния:

$$\mathbf{x}(0) = \begin{bmatrix}-3\\-2\end{bmatrix}.$$

Для алгоритма оценки были выбраны следующие параметры:

$$\alpha = 1000, \mathrm{M} = 10^{12}, \beta = 1, f_0 = 0{,}1.$$

На вход системы был подан синусоидальный сигнал:

$$u(t) = \sin(t).$$

На рисунке 1 приведены переходные процессы оценки неизвестных параметров $\hat{\boldsymbol{\Theta}}$. На рисунках 2…9 приведены переходные процессы по ошибке оценивания по каждому из неизвестных параметров. На рисунках 10 и 11 приведены переходные процессы по ошибке оценки компонент вектора состояния исходной нелинейной нестационарной системы.

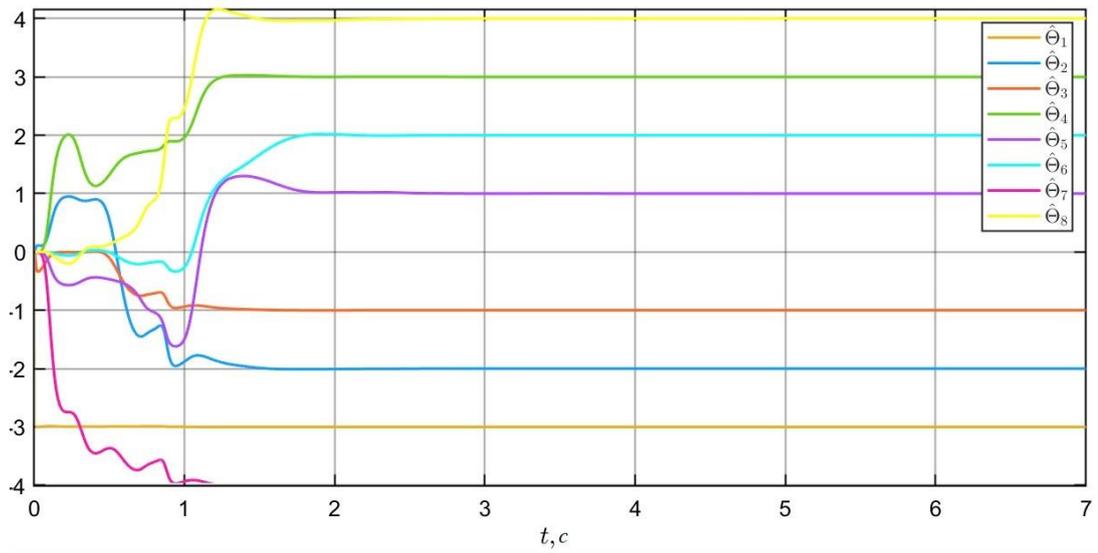

Рисунок 1 – Переходные процессы по оценке неизвестных параметров

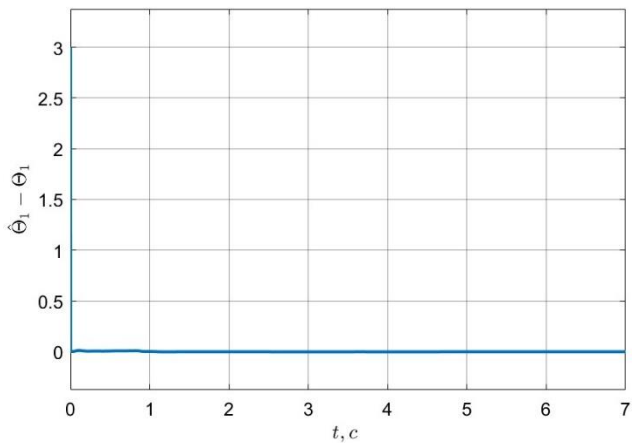

Рисунок 2 – Переходной процесс по ошибке $\hat{\Theta}_1 - \Theta_1$

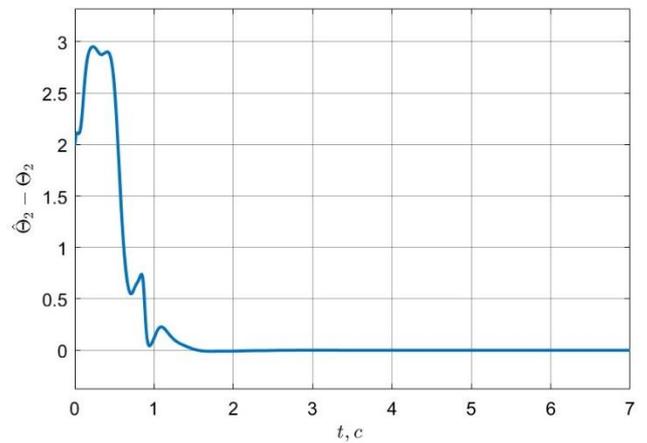

Рисунок 3 – Переходной процесс по ошибке $\hat{\Theta}_2 - \Theta_2$

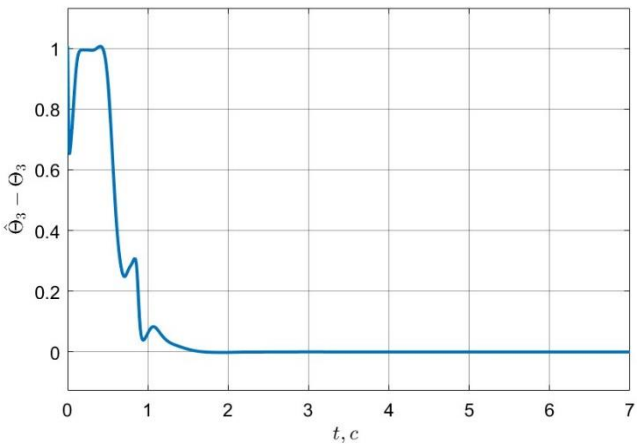

Рисунок 4 – Переходной процесс по ошибке $\hat{\Theta}_3 - \Theta_3$

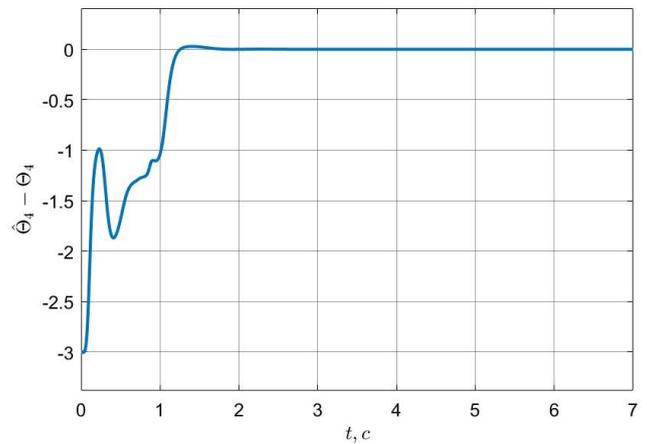

Рисунок 5 – Переходной процесс по ошибке $\hat{\Theta}_4 - \Theta_4$

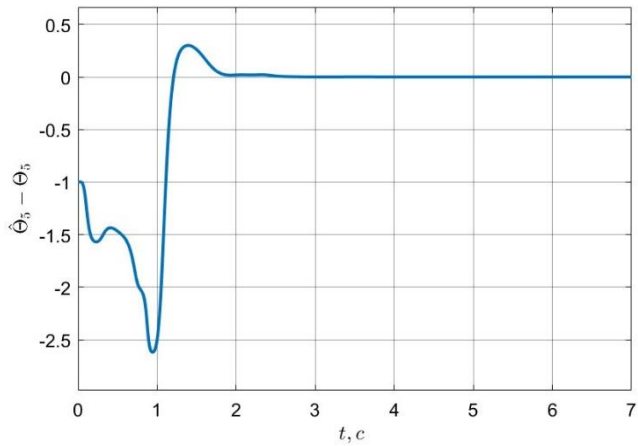

Рисунок 6 – Переходной процесс по ошибке $\widehat{\Theta}_5 - \Theta_5$

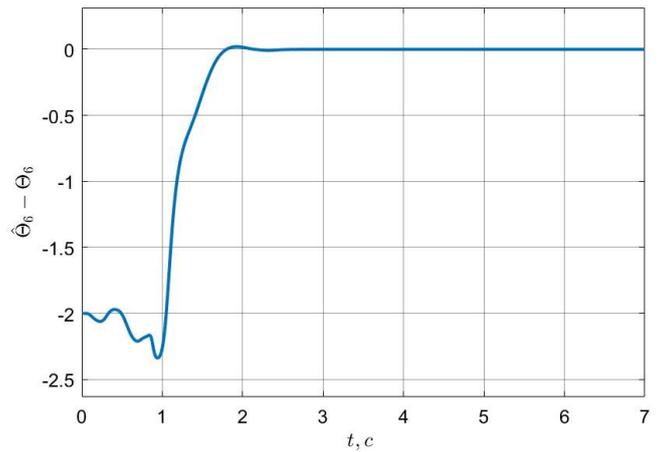

Рисунок 7 – Переходной процесс по ошибке $\widehat{\Theta}_6 - \Theta_6$

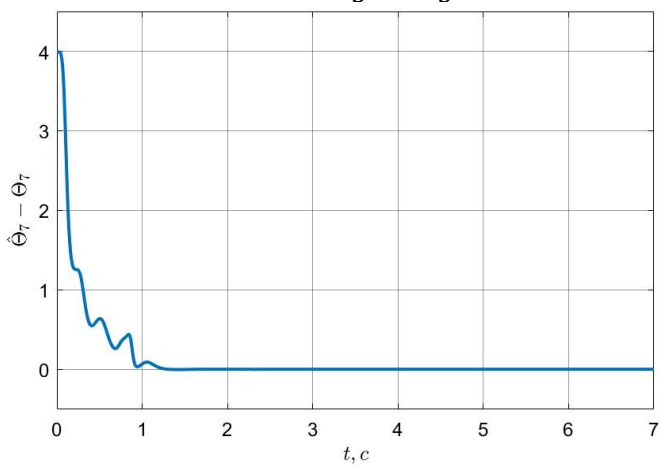

Рисунок 8 – Переходной процесс по ошибке $\widehat{\Theta}_7 - \Theta_7$

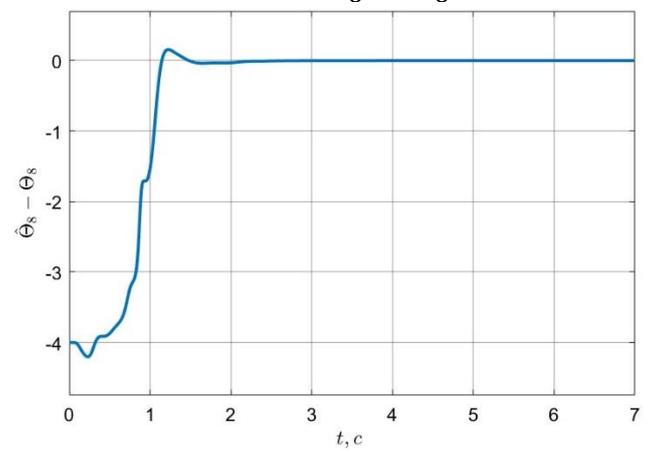

Рисунок 9 – Переходной процесс по ошибке $\widehat{\Theta}_8 - \Theta_8$

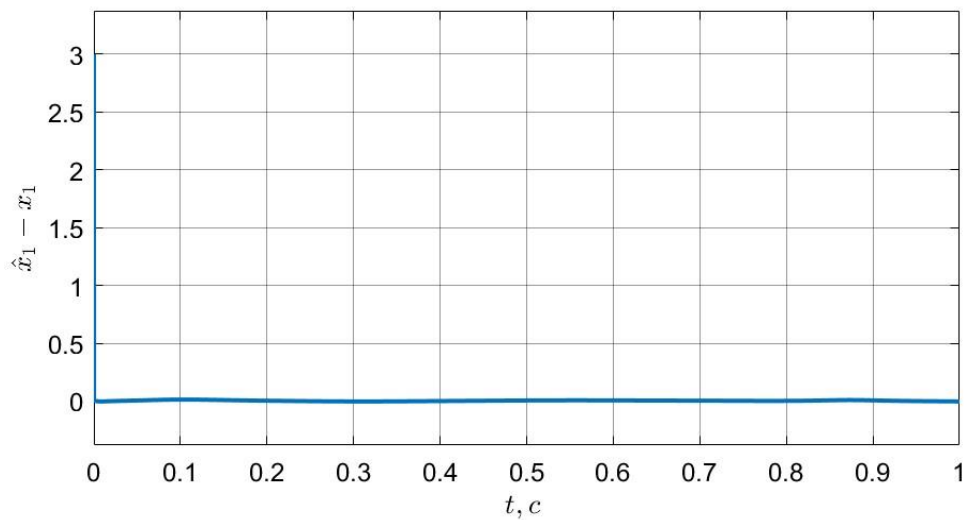

Рисунок 10 – Переходной процесс по ошибке $\hat{x}_1 - x_1$

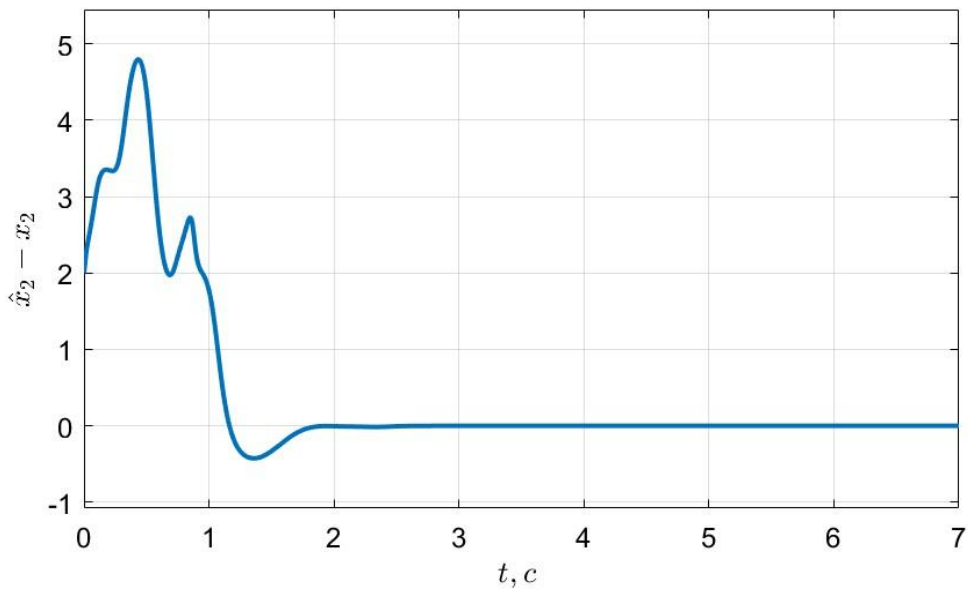

Рисунок 11 – Переходной процесс по ошибке $\hat{x}_2 - x_2$

Целью работы была разработка наблюдателя состояния системы (1), обеспечивающего сходимость оценки неизвестных параметров и вектора состояния к реальным их значениям, что означает сходимость ошибки оценивания к нулю. На представленных графиках видно, что ошибки оценивания всех параметров, а также компонент вектора состояния, сходятся к нулю, что полностью соответствует поставленной цели.

Результаты моделирования, представленные на рисунках 1…11, демонстрируют работоспособность предложенного наблюдателя переменных состояния нелинейной нестационарной системы (1), содержащей неизвестные постоянные параметры.

## Заключение

В работе предложен адаптивный наблюдатель состояния нелинейной нестационарной системы, параметры которой частично неизвестны. Неизвестные постоянные параметры содержатся в матрице (векторе) управления, а также в нелинейном компоненте. Измеряется только выходная переменная, входной сигнал полагается известным. Работоспособность предложенного алгоритма проиллюстрирована математическим моделированием.

В дальнейшем планируется развитие результатов, полученных в данной статье. В частности, дальнейшие исследования будут направлены на разработку наблюдателей состояния нелинейных систем, содержащих переменные неизвестные параметры, а также функционирующих при наличии запаздывания.

Ожидаемые результаты позволят решать прикладные проблемы синтеза наблюдателей переменных состояния для различного класса сложных технических систем, таких как импульсные преобразователи напряжения, химико-биологические реакторы и энергетические системы.

## Список литературы

1. *Ortega R., Bobtsov A., Nikolaev N., Schiffer J., Dochain D.* Generalized parameter estimation-based observers: Application to power systems and chemical–biological reactors // Automatica. 2021. Vol. 129. P. 109635

2. *Ljung L.* System identification // Signal analysis and prediction. – Birkhäuser, Boston, MA, 1998. P. 163-173

3. *Sastry S., Bodson M.* Adaptive Control: Stability, Convergence and Robustness, Prentice-Hall, New Jersey, 1989

4. *Бобцов А.А., Николаев Н.А., Ортега Мартинес Р., Слита О.В., Козачёк О.А.* Адаптивный наблюдатель переменных состояния линейной нестационарной системы с частично неизвестными параметрами матрицы состояния и вектора входа // Мехатроника, автоматизация, управление. 2022. Т. 23. № 6. С. 283-288

5. *Каленова В. И.* Линейные нестационарные системы и их приложения к задачам механики: учебное пособие / *В. И. Каленова, В. М. Морозов.* — Москва: ФИЗМАТЛИТ, 2010. — 208 с. — ISBN 978-5-9221-1231-4

6. *Баландин Д. В., Коган М. М.* Управление и оценивание в линейных нестационарных системах на основе эллипсоидальных множеств достижимости // Автоматика и телемеханика, 2020, № 8, С. 8–28.

7. *Haotian Xu, Shuai Liu, Shangwei Zhao, Jingcheng Wang,* Distributed control for a class of nonlinear systems based on distributed high-gain observer // ISA Transactions, 2023

8. *Sunjeev Venkateswaran, Costas Kravaris,* Linear Unknown Input Observers for Sensor Fault Estimation in Nonlinear Systems // IFAC-PapersOnLine, Volume 56, Issue 1, 2023, p. 61-66

9. *Bobtsov A., Ortega R., Yi B., Nikolaev N.* Adaptive state estimation of state-affine systems with unknown time-varying parameters // International Journal of Control, 2021, p. 1-13

10. *Glushchenko A., Lastochkin K.* Robust Time-Varying Parameters Estimation Based on I-DREM Procedure // arXiv preprint arXiv:2111.11716, 2021

11. *Gao F., Jiang G., Zhang Z., Song J.* An adaptive observer for actuator and sensor fault diagnosis in linear time-varying systems // Proceedings of the 10th World Congress on Intelligent Control and Automation. IEEE, 2012, p. 3281-3285

12. *Wang F., Zong M., Chen W.* Fault diagnosis of linear time-varying system based on high gain adaptive compensation sliding mode observer // 2017 IEEE 2nd Information Technology, Networking, Electronic and Automation Control Conference (ITNEC). IEEE, 2017, p. 1688-1691

13. *Кочетков С. А.* Об одном алгоритме идентификации параметров в линейных нестационарных системах // Труды IX международной конференции "Идентификация систем и задачи управления" SICPRO'12, 2012, С. 195-209

14. *Bobtsov A., Nikolaev N., Slita O., Kozachek O., Oskina O.* Adaptive observer for a LTV system with partially unknown state matrix and delayed measurements//14th International Congress on Ultra Modern Telecommunications and Control Systems and Workshops, ICUMT 2022, 2022, p. 165-170

15. *Rugh W. J.* Linear system theory. Prentice-Hall, Inc., 1996

16. *Tranninger M., Seeber R., Zhuk S., Steinberger M. and Horn M.* Detectability Analysis and Observer Design for Linear Time Varying Systems. IEEE Control Systems Letters. 2020, Vol. 4, № 2, p. 331-336

17. *Tranninger M., Zhuk S., Steinberger M., Fridman L., Horn M.* Non-Uniform Stability, Detectability, and, Sliding Mode Observer Design for Time Varying Systems with Unknown Inputs //arXiv preprint arXiv:1809.06460, 2018

18. *Aranovskiy, S., Bobtsov, A., Ortega, R., and Pyrkin, A*. Performance enhancement of parameter estimators via dynamic regressor extension and mixing. IEEE Transactions on Automatic Control, 2016, Vol. 62, № 7, p. 3546-3550

19. *Aranovskiy S., Bobtsov A., Ortega R., Pyrkin. A*. Parameters estimation via dynamic regressor extension and mixing // 2016 American Control Conference (ACC). IEEE, 2016, p. 6971-6976.

20. *Мирошник И. В., Никифоров В. О., Фрадков А. Л.* Нелинейное и адаптивное управление сложными динамическими системами, 2000.